\documentclass[12pt,preprint,flushrt]{aastex}
\doublespace
\usepackage{graphicx}
\begin{document}

\title{Bounds on Compactness for LMXB Neutron Stars from X-ray Burst 
Oscillations}
\author{Nitya R. Nath\altaffilmark{1,2}, Tod E. Strohmayer \& Jean H. Swank}
\affil{Laboratory for High Energy Astrophysics, NASA's Goddard Space Flight 
Center, Greenbelt, MD 20771; stroh@clarence.gsfc.nasa.gov}
\altaffiltext{1}{Raytheon ITSS, Lanham, MD 20706}
\altaffiltext{2}{Currently with: Science Systems and Applications, Lanham, MD 
20706}
\email{stroh@clarence.gsfc.nasa.gov}

\begin{abstract}

We have modelled X-ray burst oscillations observed with the Rossi X-ray Timing 
Explorer (RXTE) from two low mass X-ray binaries (LMXB): 4U 1636-53 with a  
frequency of 580 Hz, and 4U 1728-34 at a frequency of 363 Hz. We have computed
least squares fits to the oscillations observed during the rising phase of 
bursts using a model which includes emission from either a single circular hot 
spot or a pair of circular antipodal hot spots on the surface of a neutron 
star. We model the spreading of the thermonuclear hot spots by assuming that 
the hot spot angular size grows linearly with time. We calculate the flux as a 
function of rotational phase from the hot spots and take into account photon 
deflection in the relativistic gravitational field of the neutron star 
assuming the exterior space-time is the Schwarzschild metric. We find 
acceptable fits with our model in a $\chi^2$ sense, and we use these to place 
constraints on the compactness of the neutron stars in these sources. For 
4U 1636-53, in which detection of a 290 Hz sub-harmonic supports the two spot 
model, we find that the compactness (i.e., mass/radius ratio) is constrained 
to be $M/R < 0.163$ at 90 \% confidence ($G = c = 1$). This requires a 
relatively stiff equation of state (EOS) for the stellar interior. 
For example, if the neutron star has a mass of $1.4 M_{\odot}$ then its radius 
must be $> 12.8$ km. Fits using a single hot spot model are not as highly 
constraining. We discuss the implications of our findings for recent efforts 
to calculate the EOS of dense nucleon matter and the structure of neutron 
stars. 

\end{abstract}


\newpage

\section{Introduction}

X-ray brightness oscillations with frequencies in the 300 - 600 Hz range have 
now been observed during thermonuclear X-ray bursts from 10 LMXB systems 
(see Strohmayer 2001 for a recent review).  Substantial evidence suggests 
that rotational modulation of a localized hot spot or a pair of antipodal 
spots is responsible for the observed oscillations, especially during the 
rising phase (see for example Strohmayer, Zhang \& Swank 1997; Heise 2000). 
As the mass to radius ratio, $M/R$ or ``compactness'', of a neutron star 
increases, the deflection of photons by its relativistic gravitational field 
becomes stronger and consequently a greater fraction of the stellar surface
is visible to an observer at any given time. This effect weakens the spin 
modulation pulsations produced by a rotating hot spot on the neutron star
surface. Because of this effect, Strohmayer et al. (1997) suggested that 
modelling of the burst oscillation amplitude could in principle provide a 
constraint on the neutron star compactness. Strohmayer, Zhang \& Swank (1997) 
investigated the temporal evolution of the amplitude of burst oscillations from
4U 1728-34 and showed that a simple model of an expanding hot spot on a neutron
star was in qualitative agreement with the data. 
Miller \& Lamb (1998) performed a study of the dependence of the 
oscillation amplitude from a point-like hot spot on the stellar compactness, 
the surface rotational velocity, and the spectrum of the surface emission, 
and showed that if two antipodal spots are present, the resulting limits on 
the compactness can be highly constraining. Weinberg, Miller, \& Lamb (2000) 
have recently performed similar calculations but allow for hot spots of finite 
size. Psaltis, Ozel, \& DeDeo (2001) have also recently investigated the
effects of relativistic photon deflection on the inferred properties of 
thermally emitting neutron stars. 

Miller (1999) reported the detection of a 290 Hz sub-harmonic of the stronger
580 Hz oscillation frequency in a study of 5 bursts from 4U 1636-53. This led 
him to suggest that the neutron star spin frequency is actually 290 Hz in this
source and that two antipodal hot spots produce the 580 Hz modulation. The 
observation of a pair of high frequency quasi-periodic oscillations (QPO) with 
a frequency separation of $\sim 251$ Hz in this source (Mendez, van der Klis, 
\& van Paradijs 1998), has also been interpreted, in the context of a beat 
frequency model for the high frequency QPO, as evidence for a neutron star spin
frequency of $\sim 290$ Hz rather than 580 Hz (see Miller, Lamb \& Psaltis 
1998). We note, however, that recent efforts to confirm the sub-harmonic 
detection in subsequent bursts from 4U 1636-53 have not been successful 
(Strohmayer 2001). 

Strohmayer et al. (1998a) reported very large amplitude oscillations at 580 Hz 
during the rising phase of some bursts from 4U 1636-53. This combination of 
large measured amplitudes near burst onset and the evidence that two hot spots 
may produce the modulation, make 4U 1636-53 perhaps the best source currently 
known in which to constrain the neutron star mass and radius based on the 
properties of burst oscillations.  Here we report on our efforts to 
do this by detailed modelling of the burst oscillations observed during the 
rising phase of bursts. We focus on 4U 1636-53 because if the two hot spot 
conjecture is correct for this object then our results place strong constraints
on the neutron star compactness. However, we also summarize our results for 
4U 1728-34, a source which has also shown strong oscillations during the rising
phase of bursts. The plan of this paper is as follows. In \S 2 we discuss the
basic features and assumptions of our model. In \S 3 we outline the method of
calculation. In \S 4 we describe our model fitting procedures and our results 
for both single and antipodal hot spot models. We also summarize the results 
of fits to data from 4U 1636-53 and 4U 1728-34. In \S 5 we summarize our 
results and discuss them in the context of recent efforts to constrain the EOS
of neutron star matter. We also discuss future steps we will take to improve 
the hot spot model.

\section{Model Assumptions}

Both spectral and temporal evidence indicate that the X-ray emission near the
onset of at least some thermonuclear bursts is localized to a ``hot spot'' 
which spreads in some fashion until eventually encompassing all of the neutron
star surface (see for example Strohmayer, Zhang \& Swank 1997). This likelihood
was also recognized early on in theoretical studies of thermonuclear bursts 
(Joss 1978). Motivated by this we model the burst rise by assuming that all the
burst emission comes from either one or a pair of circular hot spots which 
expand linearly in angular size with time. The rest of the neutron star surface
is assumed dark. Photon trajectories are computed assuming the Schwarzschild 
metric describes the space-time exterior to the star. This is a reasonable 
approximation since the influence of the neutron star's rotation on the 
space-time only affects the oscillation amplitude to second order (Miller \& 
Lamb 1996). For the present work we shall only investigate bolometric 
modulations across the full $\sim 2 - 90$ keV bandpass of the RXTE Proportional
Counter Array (PCA). We shall also ignore Doppler shifts and relativistic 
aberration produced by the rotational motion of the hot spot (see for example 
Miller 1999; Chen \& Shaham 1989). We discuss later the likely influence on our
results of this approximation. 

Our model is uniquely characterized by seven parameters: (1) an overall source 
intensity or normalization, $S$, which can be thought of as the flux leaving 
unit surface area of the neutron star. (2) neutron star compactness, $\beta = 
M/R$, where $M$ and $R$ are the stellar mass and radius, respectively, 
(3) initial angular size of the spot (half of the subtended angle), 
$\alpha_0$, (4) angular growth rate of the hot spot, $\dot\alpha$, (5) initial 
rotational phase, $\delta_0$, (6) latitude of the spot center, $\theta_s$, 
measured from the rotational equator, and (7) latitude of the observers line of
sight, $\theta_{obs}$, also measured from the rotational equator. 
One of our primary goals is to determine an upper bound on the compactness.  
To do this within the context of our model we set the hot spot latitude 
and observation latitude to zero. That is, both the hot spots and the line of 
sight to the observer are centered on the rotational equator. This geometry 
produces the largest possible modulation amplitude. Since any observed 
modulation must be equal to or less than this limit, and since the modulation 
amplitude decreases with increasing compactness, the upper limit follows. 
For completeness, we also investigate the influence of moving the hot spot and 
the line of sight off the rotational equator. The geometry of our model is 
illustrated in Figure 1. Related hot spot models have been worked out by 
Pechenick, Ftaclas, \& Cohen (1983) and Strohmayer (1992).

\section{Method of Calculation}

The geometry of a photon trajectory in relation to the observers line of sight
$\vec r_{obs}$ is shown in Figure 1. The figure is drawn with $\theta_s =
\theta_{obs} = 0$.  For any single point on the hot spot with radius vector 
$\vec r$, the path of a photon reaching the observer lies in the plane of 
$\vec r$ and $\vec r_{obs}$, and is asymptotically parallel to $\vec r_{obs}$  
with impact parameter $b$. The two angles, $\phi$ (between $\vec r$ 
and $\vec r_{obs}$) and $\psi$ (the emission angle with respect to the 
surface normal), complete the description. For non-zero $\theta_s$ and 
$\theta_{obs}$, the deflection geometry remains the same, only the plane in 
which the desired trajectory lies (the plane of $\vec r$ and $\vec r_{obs}$) 
changes.  The angle $\phi$ can be expressed as,
\begin{equation}
\phi = \int_0^{\sin^{-1}(\hat b)} \left [ 1 - 2(M/R)(1-\sin^3 y / \hat b ) / 
(1 - \sin^2 y) \right ] ^{-1/2} dy,
\end{equation}
where $\hat b = b/b_{max}$ is the reduced impact parameter, $b_{max} = 
R(1-2(M/R))^{-1/2}$, and $M$ and $R$ are the stellar mass and radius
respectively. This form for the angle $\phi$ is somewhat non-standard compared 
to previous work. More commonly $\phi$ is expressed as
\begin{equation}
\phi = \int_0^{M/R} \left [u_b^2 - (1-2u)u^2 \right ] ^{-1/2} du,
\end{equation}
where $u_b \equiv M/b$ (see for example, Pechenick, Ftaclas \& Cohen 1983; 
Miller \& Lamb 1998). Our rationale for rewriting the integrand is twofold; 
first, to explicitly show what parameters $\phi$ depends on, and second to 
remove singular behavior of the integrand to facilitate numerical evaluation 
of the integral. Changing variables in (2) to $u = (M/R) x$ results in the 
following expression;
\begin{equation}
\phi = \int_0^1 \left [\hat b ^{-2} (1-\frac{2M}{R}) - 
(1-\frac{2M}{R}x)x^2 \right ] ^{-1/2} dx.
\end{equation}
As $M/R$ becomes small this integral has the form,
\begin{equation}
\phi = \int_0^1 \left [\hat b ^{-2}  - x^2 \right ] ^{-1/2} dx,
\end{equation}
which has singular behavior as $\hat b$ and $x$ approach unity. The second
change of variables to $y=sin^{-1}\hat b x$ is motivated by the form of 
equation (4) above, whose solution corresponds to the inverse sine function. 
With this final substitution we arrive at the
expression in equation (1), which explicitly shows the dependence of $\phi$
on $M/R$ and $\hat b$, and is well defined and non singular. 
As $\hat b$ varies from 0 to 1, $\phi$ varies from 0 to $\phi_{max}$,
the maximum value of $\phi$, which is attained when a photon is emitted 
tangentially to the stellar surface. We note several interesting limiting
cases; for $M/R = 0$, 0.284, 0.331, 0.33333, we have $\phi_{max} = \pi/2$, 
$\pi$, $2\pi$, and $\infty$, respectively. The last case, $\phi_{max} = 
\infty$, corresponds to the bound photon orbit at $M/R = 1/3$.  

To compute the flux as a function of rotational phase we first invert 
$\phi\; (\hat b, M/R)$ numerically to obtain tables of $\hat b$ as a function 
of $\phi$ and $M/R$. We use Gaussian quadratures to solve the integral 
numerically. The method is fast and converges quickly. 
For a given $M/R$ and each $\phi = \cos^{-1} (\vec r 
\cdot \vec r_{obs})$ we then find $\hat b$ and compute $\cos\psi = 
(1 - \hat b^2 )^{1/2}$.
The observed flux is then given by $\int I_{\nu} \cos\psi\; d\Omega$, where 
$I_{\nu}$ is the local specific intensity at the surface of the neutron star, 
and the integral is carried out over the hot spot or spots. For the specific 
intensity we use both an isotropic emission function, $I_{\nu} = 1$ and an 
angular dependent beaming function consistent with emission from a grey 
scattering atmosphere, $I_{\nu} = 3/5 \cos\psi + 2/5$ (see Chandrasekhar 1960).
Such a function should be appropriate for bursting neutron star atmospheres 
which are dominated by Thomson scattering (London, Taam, \& Howard 1986). 
Figure 2 shows several examples of light-curves computed with our model using 
one hot spot and different values of $M/R$. The decrease in modulation 
amplitude with increasing compactness is clearly evident. 

\section{Data Analysis Procedures and Results}

We searched the available RXTE data from 4U 1636-53 and 4U 1728-34 for bursts 
and selected for analysis four from 4U 1636-53 and two from 4U 1728-34 which 
showed particularly strong oscillations during the rising phase. The data are
in the form of X-ray event times recorded with 125 $\mu$s resolution across
the full 2 - 90 keV PCA bandpass.
In order to fit our model we first break up the rising interval from each burst
into a number, $n_{interval}$, of contiguous subintervals. Within each 
subinterval we epoch fold the data into $n_{bin}$ phase bins using the 
oscillation frequency determined from a power spectral analysis of the entire 
rising interval. We then perform a $\chi^2$ minimization by computing 
$\chi^2 = \sum_{i=1}^N (O_i - M_i)^2 / \sigma^2 $. Here $O_i$  and $M_i$ are 
the numbers of observed and predicted counts, respectively, in the 
$i$$^{\rm th}$ data bin. For $\sigma^2$ we use the Poisson variance, which is
simply equal to the number of counts in the bin. In general we also add a 
constant background level to the model as a way of modeling the pre-burst, 
accretion driven flux, which we assume is not associated with the burst. This 
also implies a tacit assumption that the accretion driven flux is not 
significantly altered by the burst. This quantity is well determined by the 
pre-burst data, so typically we do not treat it as a model parameter. In 
general the total length of data that we fit does not extend all the way to
the peak of the burst for a number of reasons. The oscillation has usually 
dropped below our detection threshold before the peak is reached and often 
episodes of radius expansion also begin before the count rate reaches a 
maximum. In general, our assumptions regarding the growth of the hot spot
should be most valid the closer we remain to the onset of the burst. This also
tends to maximize our signal to noise ratio in data from a given burst since
the modulation amplitude is largest near burst onset.

We minimize $\chi^2$ using the Marquardt-Levenberg method and we can 
simultaneously vary all seven model parameters. Our choice regarding the number
of data bins is a tradeoff between having sufficient counts in each bin and the
need to have enough time resolution to adequately model the rise of the burst 
and hence constrain the hot spot spreading speed, $\dot\alpha$. In general we 
found that $n_{region} = 8$ and $n_{bin} = 8$ gave the best results. With this 
choice we have a total of 64 data bins. We also restrict $M/R \le 0.284$, the 
limit beyond which photons from a given point on the stellar surface can reach 
the observer along more than one unique path. In general we find acceptable 
fits using both one and two hot spots for both sources. In the remainder we 
will summarize our results and discuss the implications for neutron star 
compactness, concentrating on the two spot fits for 4U 1636-53 for the reasons 
outlined above.

\subsection{Antipodal Hot Spot Models}

Our best fitting models for bursts from 4U 1636-53 using two antipodal hot 
spots and the grey atmosphere intensity function are summarized in Table 1, 
where for each burst we give the observation date, the length of the time 
interval in which we fit the data, the best fitting model parameters and the 
minimum $\chi^2$. For these fits we have fixed to zero both the spot latitude, 
$\theta_s$ and the observers latitude, $\theta_{obs}$, and we used 64 data 
bins. With 5 free parameters we therefore have 59 degrees of freedom. Our 
minimum $\chi^2$ values are all statistically acceptable, indicating that the 
simple rotating hot spot model is consistent with the data. In Figure 3 we 
show the two spot fits for each of the four bursts from 4U 1636-53. Each panel 
shows the count rate in the PCA for the rising interval of a burst. The bursts 
are labelled by date. The vertical dashed lines denote the region in which we 
fit our model. The solid curve shows the best fitting model {\it extrapolated} 
to the time at which the entire surface of the neutron star is covered by the 
hot spots. The time resolution in these plots is not sufficient to 
resolve the oscillations, rather, this figure is meant to give the reader an 
assessment of how well the model does in describing the gross time evolution of
each burst. There are several things to note from Figure 3. First, the fits 
{\it within} each interval are quite good, and they also extrapolate beyond 
the fitting interval rather well over a limited portion of the burst rise.
The deviations at later times are not unexpected since in several of these 
bursts episodes of photospheric radius expansion begin at about the same time
as the model begins to deviate from the burst rise. Indeed the burst on 
08/20/98 did not show radius expansion and in this case the model extrapolates
rather well for most of the rise. All the other bursts show radius expansion 
near the time that the model deviates from the data. Second, the maximum count 
rates inferred from our model for bursts 12/28/96 and 08/19/98 are quite 
similar. Since these bursts were quite similar in their peak fluxes, the model 
normalizations, which can be thought of as an averaged description of the 
thermonuclear burning, should also be similar and indeed they are. Note that
though these two bursts have similar peak fluxes they do not have similar rise
times, and our model succesfully accounts for this difference. 
Although the models are clearly inadequate to describe the details of the 
{\it entire} burst rise, they do better the closer one stays to the burst 
onset, and this behavior is the most relevant with regard to fitting the 
oscillations and constraining $M/R$.

Figure 4 graphically illustrates how well the model can fit the observed 
oscillations by comparing the best fit model and data for several different 
fits to the 12/28/96 burst from 4U 1636-53. Shown are the best fitting two 
spot model with $\theta_s = \theta_{obs} = 0$ (solid); the best fitting one 
spot model with $\theta_s = \theta_{obs} = 0$ (dashed); and the best fitting 
one spot model with all parameters free to vary (dotted). Since fits to the 
other bursts all look very similar we did not feel it was essential to show 
similar plots for each individual burst.  

The derived best-fit compactness for the four bursts from 4U 1636-53 span a 
rather tight range from $\beta = 0.075$ to $0.134$. In Figure 5 we show the 
best-fit values of $M/R$ and their uncertainties. We fit a constant, 
$\beta_{avg}$, to the four values and find they are consistent with a single
value for the compactness of $\beta_{avg} = 0.126$ (solid horizontal line in 
Figure 5). The $\chi^2$ per degree of freedom for the fit was 0.2.  
In order to derive a firm upper limit on the compactness we investigated
the confidence region for $\beta_{avg}$ and found the values 
of $\beta$ which increased $\chi^2$ by 2.71 (for $90\%$ confidence) and 6.63 
(for $99\%$ confidence). These values are also shown in Figure 5 as the 
dashed ($90\%$) and dot-dashed ($99\%$) horizontal lines in Figure 5. The 
derived upper limits are $\beta_{90\%} = 0.163$ and $\beta_{99\%} = 0.183$.

Since we do not in general know the orientation of the binary systems in 
which the neutron stars reside we performed the fitting under the assumption 
that the hot spot and observer are both in the plane of the rotational equator.
This geometry gives the maximum rotational modulation. Thus each fitted value 
for $\beta$ from a different burst gives a measure of the maximum allowable
compactness of the neutron star. However, each individual measurement has 
associated with it a rather large uncertainty. Thus, our methodology in 
deriving an upper limit on $\beta$ is to combine a number of these independent 
measurements in order to reduce the overall uncertainty. In this way 
$\beta_{avg}$ is our best estimate of how large the compactness of the neutron 
star is, but this estimate too is not exact and has a confidence region 
associated with it. It is the confidence region on $\beta_{avg}$ that we use 
to determine a final upper limit. This may not be a unique statistical 
methodology, but we feel it is reasonable given the nature of the other model 
assumptions we have made. We will discuss the implications of our compactness 
limits for the neutron star EOS in the next section. 

We also computed fits allowing the two angles $\theta_s$ and $\theta_{obs}$
to vary. As might be expected we find the inclusion of the additional 
parameters improves the fits, but only marginally. With these parameters free
to vary we find that $M/R$ tends to decrease, and both $\theta_s$ and 
$\theta_{obs}$ move off the rotational equator. We find, however, no 
stationary solutions in $M/R$ with all seven parameters varying. 
These results serve to illustrate the basic correlation between compactness and
the hot spot and viewing geometries. If the spot moves or is viewed away
from the rotational equator then the inferred value of $M/R$ must decrease in 
order to make up for the loss of modulation amplitude produced by a less than 
favorable geometry.  Since realistic neutron star EOSs cannot support stars
with arbitrarily small $M/R$, if the two spot model is correct, then our 
results suggest that the hot spots must be relatively near the rotational 
equator in order to achieve the high observed amplitudes. 
If the hot spots are linked to the poles of a magnetic field in 4U 1636-53 
(see Miller 1999), then this would suggest that the magnetic axis would have 
to be nearly perpendicular to the rotation axis.  If we assume the surface 
emission is isotropic, the fits for all the bursts are very similar, but the 
$M/R$ values are systematically lower, with the weighted mean dropping to 
$M/R = 0.05$. This is as expected, since isotropic emission produces a 
lower amplitude than the grey atmosphere beaming function.

Our results for 4U 1728-34 are quite similar to those derived for 4U 1636-53.  
The results of the two spot fits for bursts from 4U 1728-34, with $\theta_s$ 
and $\theta_{obs}$ fixed at zero and with beamed emission are also shown in 
Table 1. The weighted average of the two fits yields the value $M/R = 0.121$,
with $90\%$ and $99\%$ confidence upper limits of $0.171$ and $0.199$, 
respectively. These are similar to the limits derived for 4U 1636-53. 
Although no sub-harmonic has been detected for this source, the closeness of 
the derived $M/R$ limits for the two sources is striking, and may be an 
indication that, irrespective of the model, the actual compactness of the two 
sources is similar.  

\subsection{One Spot Models}

For one spot models we generally find there are no strong constraints on the 
compactness for either source. This results from the fact that stars even as 
compact as our computational limit, $M/R = 0.284$, can still produce a 
sufficiently large modulation amplitude to match the data. For example, the 
best fits for the 4U 1636-53 bursts with four parameters varying (ie., $M/R$ 
fixed at 0.284, $\theta_s$ and $\theta_{obs}$ fixed at zero), and with beamed 
emission, give $\chi^2$ = 69.4, 66.2, 70.6, and 75.0, for each burst 
respectively. These values are marginally higher than for the corresponding two
spot fits, however, from a statistical point of view they are still formally 
acceptable. For the one spot fits we find that $\chi^2$ monotonically decreases
as $M/R$ increases from 0 to 0.284, but never reaches a minimum. In other words
we find no meaningful upper limit to the compactness, at least within the 
confines of our model assumptions. A comparison of the $\chi^2$ values between 
the two spot and one spot fits at first glance seems to suggest that the two 
spot fits are better, however, this is misleading because the one spot fits are
not stationary in $M/R$, that is they have not converged to a minimum.

\section{Discussion and Summary}

We have shown that if two hot spots produce the observed modulation at 580 Hz 
in 4U 1636-53 then the large amplitude of oscillations near burst onset
provide a strong constraint on the compactness. In Figure 6 we show in the 
mass - radius plane our 90 and 99 \% confidence upper limits on the 
compactness $\beta=M/R$ for 4U 1636-53 from our two hot spot fits. The shaded 
region denotes the ranges of $M$ and $R$ which satisfy our compactness 
constraint and have $M > 1.4 M_{\odot}$, which we take as a reasonable 
estimate of the minimum mass of the neutron star in these old accreting 
systems. We also show several theoretical neutron star EOSs which span a range 
of stiffnesses based on current uncertainties in the exact composition 
of neutron star matter and our incomplete knowledge of the nucleon - nucleon 
interaction. Also shown in Figure 6 is our computational limit at 
$M/R \le 0.284$ (solid diagonal line).

As can be seen our results tend to favor moderately stiff to very stiff EOSs.
For example, our limits are comfortably consistent with EOS L 
(Pandharipande \& Smith 1975). However, the most recent theoretical 
calculations of neutron star EOSs which are consistent with the currently 
available nucleon scattering data are generally not as stiff as this EOS 
(see for example Akmal, Pandharipande \& Ravenhall 1998). For example, the best
EOS of Akmal, Pandharipande \& Ravenhall (1998), which is denoted APR in Figure
6, is barely consistent with our $99\%$ limit.  However,
these modern EOSs are still not rigorously self consistent, and become 
``superluminal'' (the sound speed exceeds the speed of light) above some 
density. Modifications to the EOS can be made in an ad hoc manner by setting
the sound speed equal to the speed of light above some critical or ``matching''
density (see for example Heiselberg \& Hjorth-Jensen 1999) . This has the 
effect of stiffening the EOS. Recently, Olson (2001) has 
investigated changes to the high density EOS of neutron star matter required 
by constraints derived from relativistic kinetic theory. In Figure 6 we show 
two of these modifications to the APR EOS. The thick dashed lines show the 
APR EOS modified by the kinetic theory constraints for two different matching 
densities, 0.316 fm$^{-3}$ (APR-Kin1) and 0.270 fm$^{-3}$ (APR-Kin2) 
(see Olson 2001 for a detailed discussion). With the kinetic theory 
assumptions the APR EOS is now reasonably consistent with our limits. 

Recently, Lattimer \& Prakash (2000) have argued that measurements of the
neutron star radius to about $10\%$ precision should be sufficient to usefully 
constrain the neutron star EOS. They showed that as long as extreme softening 
of the EOS does not occur in the vicinity of 
nuclear matter equilibrium density then the stellar radius is almost 
independent of the mass. Since observed neutron star masses cluster rather 
closely around 1.4 $M_{\odot}$ they argued that the more important quantity in 
terms of constraining the EOS is the stellar radius. Since the neutron stars 
in LMXBs are upwards of $10^8$ yr old and they have been accreting most of 
their lifetime, it is very likely that they are at least more massive than the 
$1.4 M_{\odot}$ typically found for younger neutron stars (Thorsett \& 
Chakrabarty 1999). If this is the case, then our results place a rather firm 
lower limit on $R$ of about 11.5 km. Such a limit is consistent with the notion
that extreme softening of the EOS, as can be produced by pion, kaon or other 
hyperon condensates, does not occur in neutron star cores (see Lattimer \& 
Prakash 2000). Since these inferences depend crucially on the two hot spot 
hypothesis for the burst oscillations from 4U 1636-53, it is vital to try and 
settle this issue in the near future. 

We have generally tried to employ the simplest assumptions consistent with 
maintaining the essential physics of the model and the observed properties of
the bursts. For this work we have neglected the Doppler shifts and relativistic
aberration produced by the rapid motion of the hot spots. Although we do not
know the rotational velocity precisely because of our uncertainty in the
stellar radius and the number of hot spots, it is likely that the velocity on 
the rotational equator is $\le 0.1\; c$. Miller \& Lamb (1998) investigated 
the effects of the rotational velocity of a point spot on the bolometric and 
energy dependent amplitude and showed that although such a velocity can have 
important effects on the amplitude measured at particular photon energies, 
they also showed that the effect on the bolometric amplitude of the rotational 
velocity is very modest (see their Figure 1d; see also Weinberg, Miller \& 
Lamb 2000). The calculations of Miller \& Lamb (1998) were for point-like spots
and hence represent upper limits to the size of any rotational effect. Since 
our model uses spots of a finite and growing angular size, the rotational 
effects, which represent an integral of the line of sight rotational velocity
over the hot spot, must be less than the estimates computed by Miller \& Lamb 
(1998). The amplitude of higher harmonics is more sensitive to the rotational 
velocity; however, the present RXTE data are not very sensitive to the shape 
of the pulses, i.e., we do not detect any higher harmonics, nor do we know of 
any published reports of significant harmonics of burst oscillations. Based on 
this and because we only investigate the bolometric amplitude we believe we are
justified in neglecting the Doppler effects for the present work. However, by 
not investigating the energy dependent effects we are indeed ignoring some 
useful information which can eventually help provide more powerful constraints 
on $M$ and $R$. We plan to improve our model by including these energy 
dependent effects and will report the results from such a study in a sequel. 

Using our model we have also begun to investigate the constraints that can be
obtained with data of a higher statistical precision than presently available.
We have found that the present RXTE data is essentially insufficient for 
constraining the hot spot and viewing geometry. However, if the count rate 
were increased by a factor of 10 - 20 times the RXTE rate then our simulations 
suggest that it will be possible to simultaneously constrain both the stellar 
compactness and the hot spot and viewing geometries. Thus future large area
timing experiments, such as the proposed Timing of Extreme X-ray 
Astrophysical Sources (TEXAS) experiment, will be extremely powerful tools for 
probing the structure of neutron stars. 

\acknowledgements

We thank Cole Miller, Craig Markwardt and Tim Olson for many helpful 
discussions and comments on the manuscript. We thank Cole Miller for 
providing some of the mass - radius relations for the equations of state 
shown in Figure 6. We also thank Tim Olson for providing the mass radius 
relations based on kinetic theory constraints to the APR equation of state. 

\vfill\eject

\newpage
\section{Figure Captions}

\figcaption[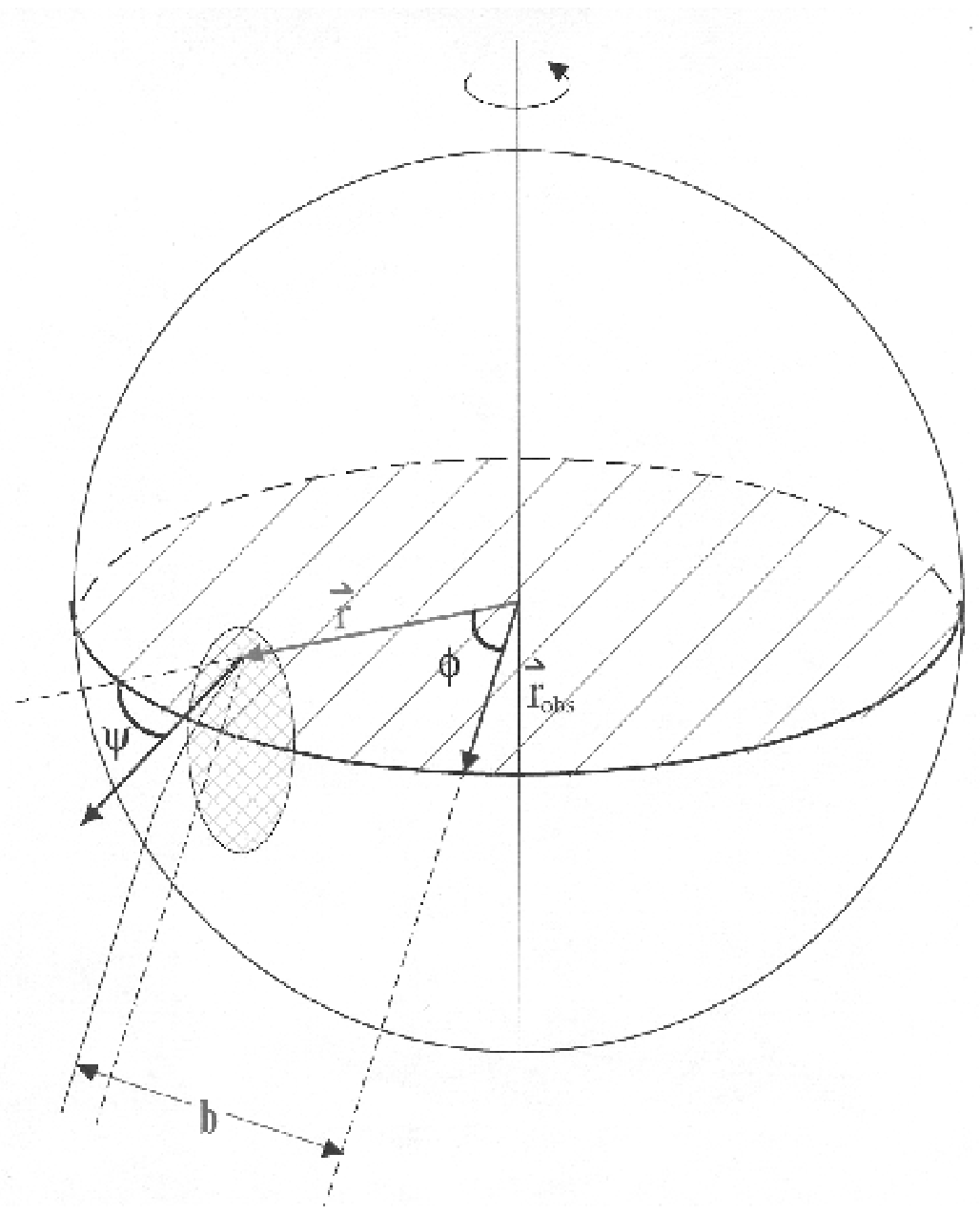]{Geometry for calculation of the flux from a hot 
spot on a rotating neutron star. Here the hot spot is situated on the 
rotational equator. See the text for a description of the relationship between 
the angles $\phi$, $\psi$ and the impact parameter, $b$. \label{fig1}} 

\vskip 10pt

\figcaption[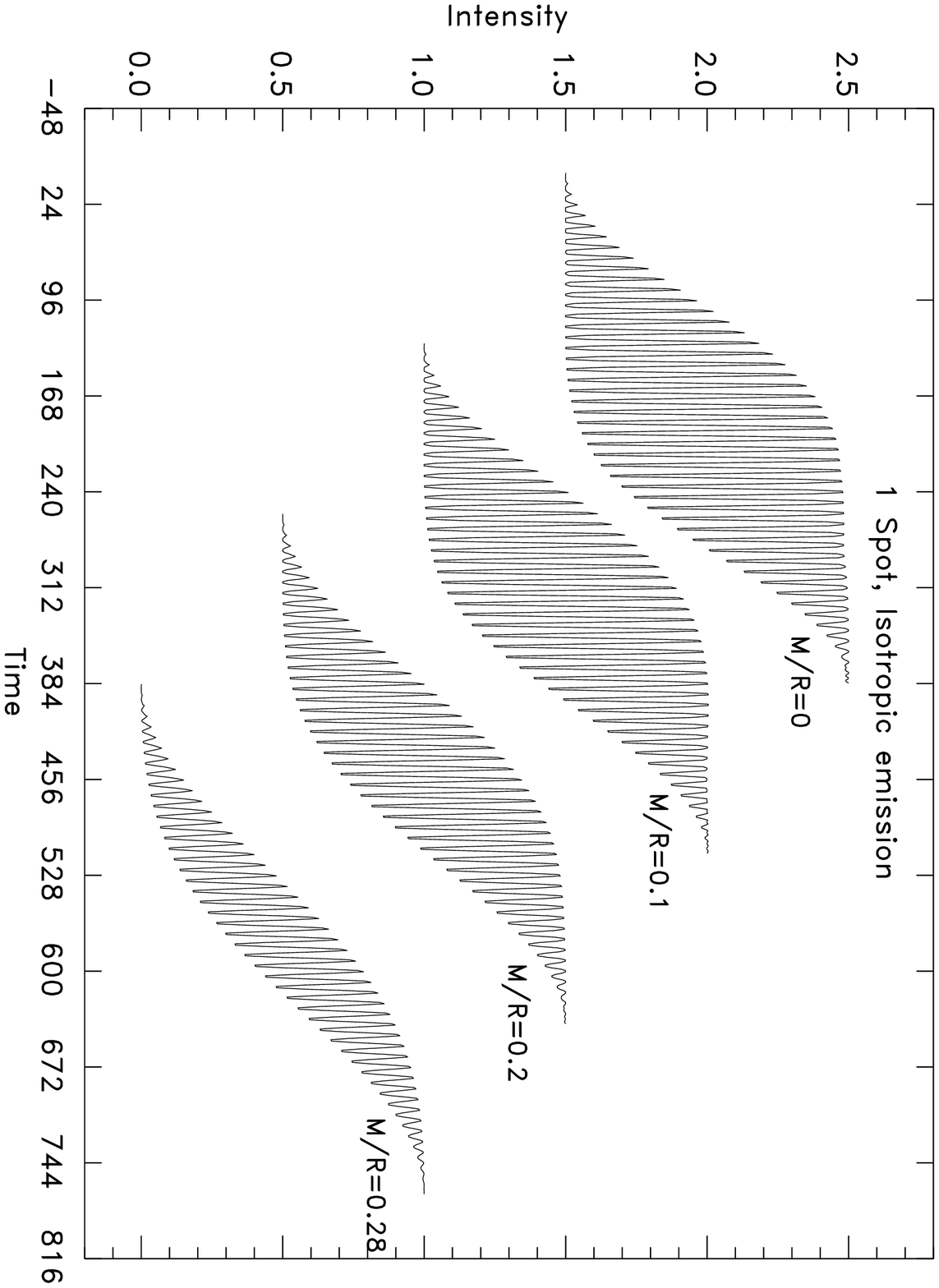]{Light-curves generated with the
rotating hot spot model for different values of the neutron star compactness.
Notice the decrease in amplitude with increasing compactness. Note also the 
decrease in amplitude and increase in flux as the hot spot spreads to encompass
the entire surface. These models were computed with one hot spot assuming
isotropic emission from the surface. The top three curves have been displaced
vertically for clarity. The qualitative behavior of the amplitude with 
compactness, $M/R$, using the grey atmosphere beaming function is the same, 
only the modulation amplitudes differ slightly.  \label{fig2}} 

\vskip 10pt

\figcaption[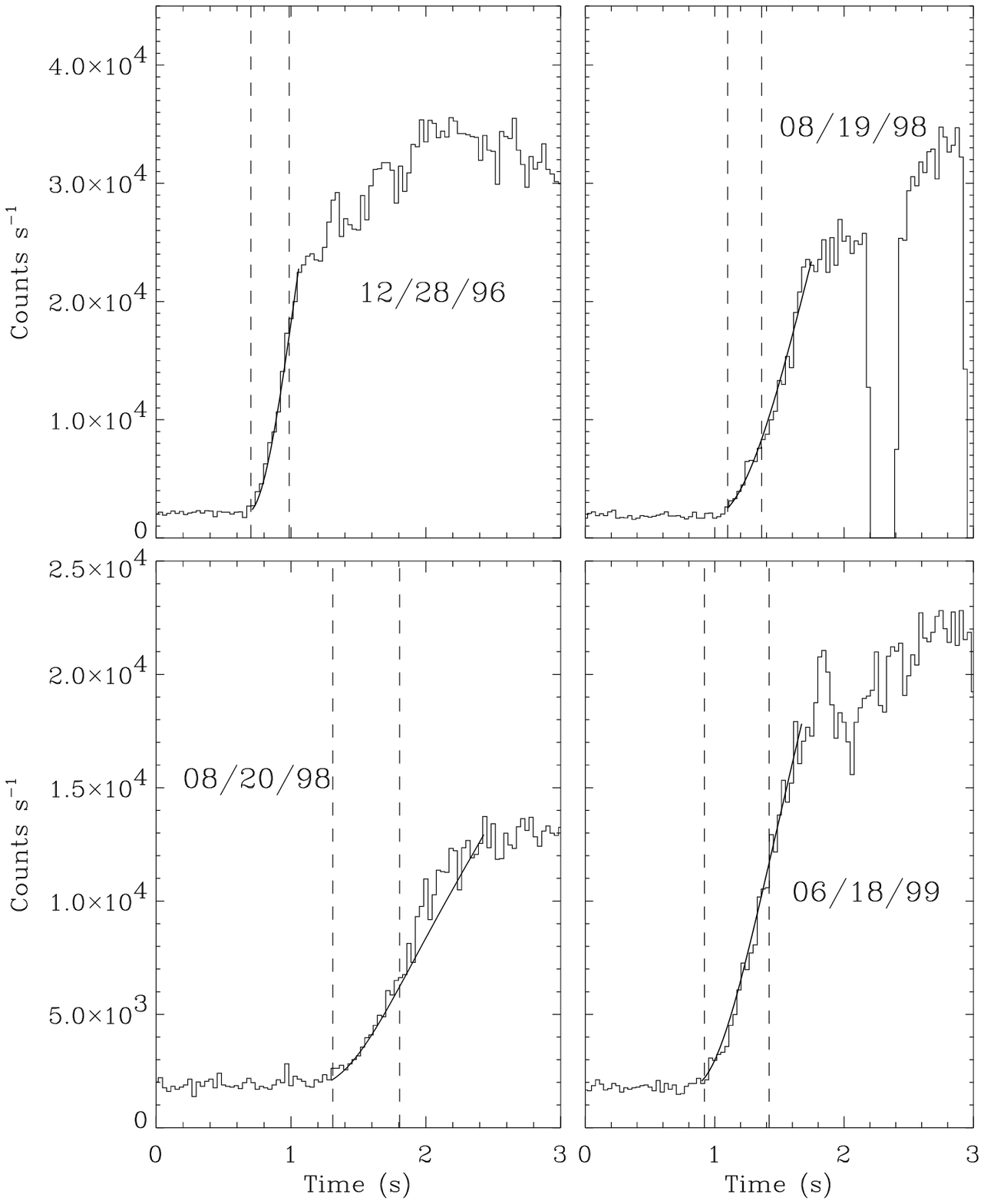]{Model fits to the bursts from 4U 1636-53. Each
panel shows the data (histogram) and model (thick solid curve) fit to the 
rising portion of a burst. The dashed vertical lines denote the time interval
in which we fit the hot spot model. The extent of each model curve covers the 
total time it takes for the hot spots to envelope the entire neutron star
surface. The bursts are labelled by date. \label{fig3}}

\vskip 10pt

\figcaption[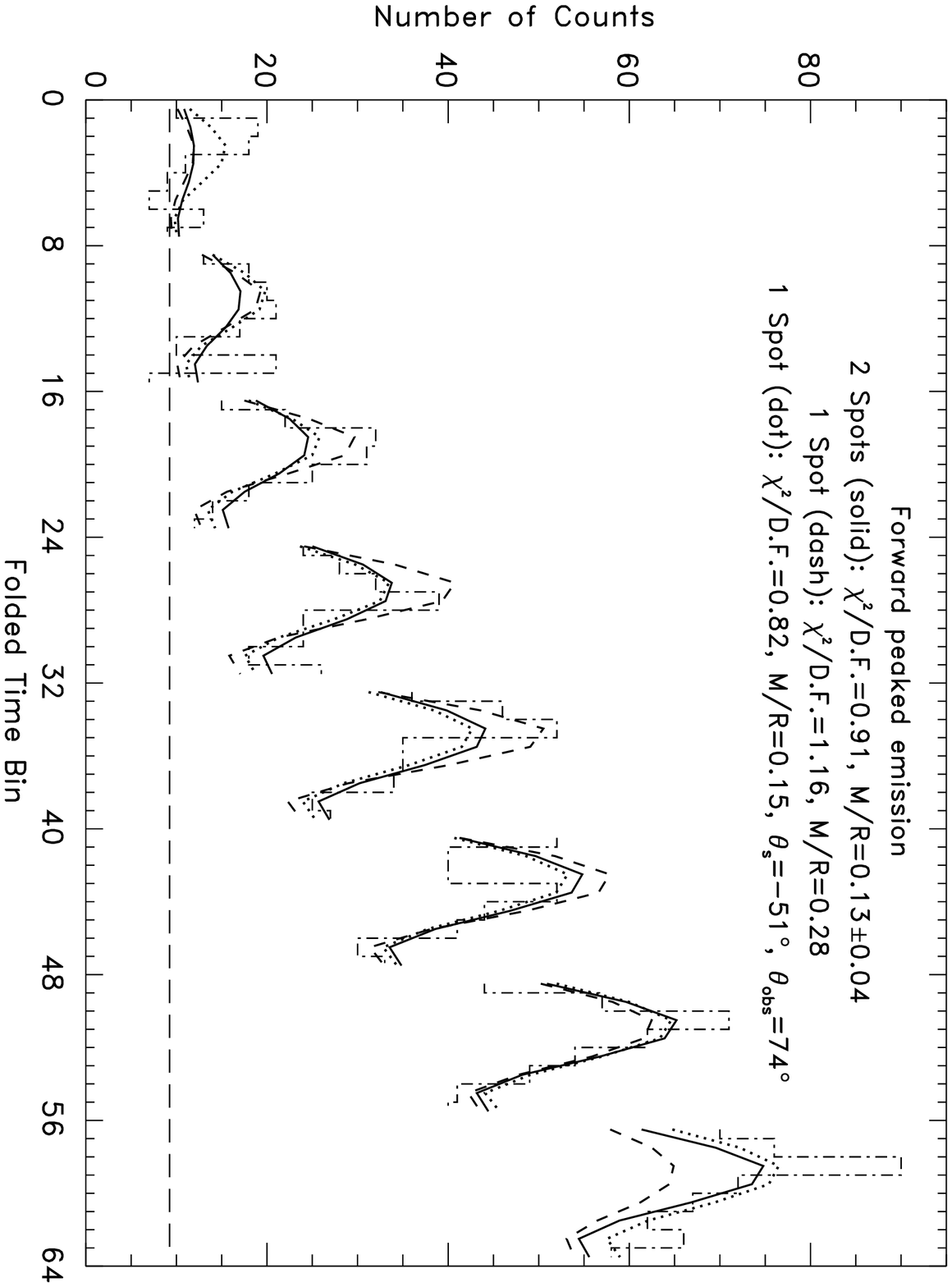]{Data and best fit models for several 
fits to the December 28th, 1996 burst from 4U 1636-53. Shown are fits using
two hot spots with $\theta_s = \theta_{obs} = 0$ (solid); one hot spot with
$M/R$ fixed at 0.284 and $\theta_s = \theta_{obs} = 0$ (dashed); and one hot 
spot with $\theta_s$ and $\theta_{obs}$ free to vary (dotted). All the fits
shown were computed with the grey atmosphere beaming function. \label{fig4}} 

\vskip 10pt

\figcaption[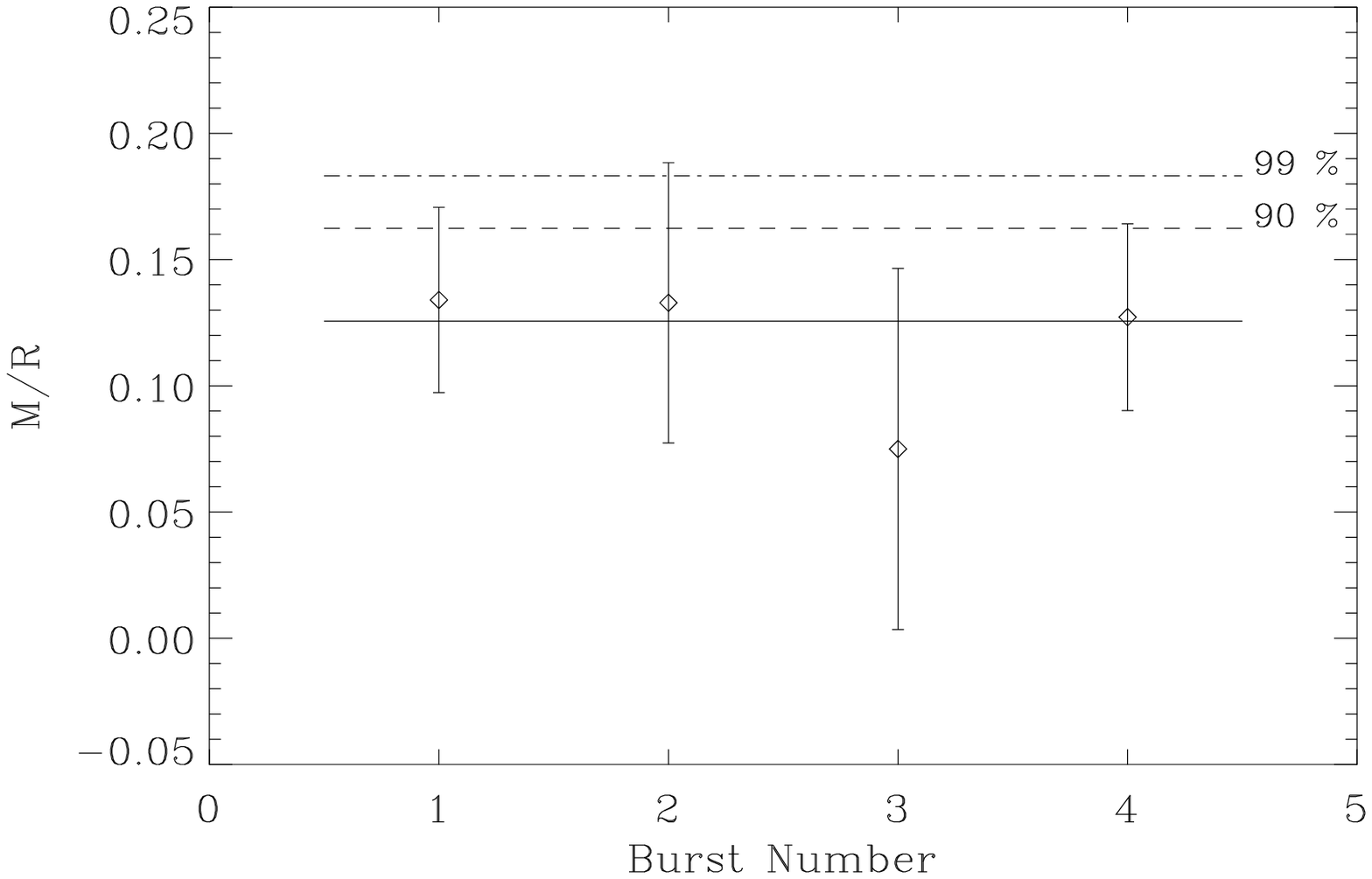]{Compactness constraints for the four bursts 
from 4U 1636-53. The solid horizontal line is the best fitting constant value 
of compactness, $\beta_{avg}$. The dashed and dot-dashed lines are the 90 and 
$99\%$ confidence upper limits on $\beta_{avg}$. The burst number corresponds
to their position in Table 1. \label{fig5}}

\vskip 10pt

\figcaption[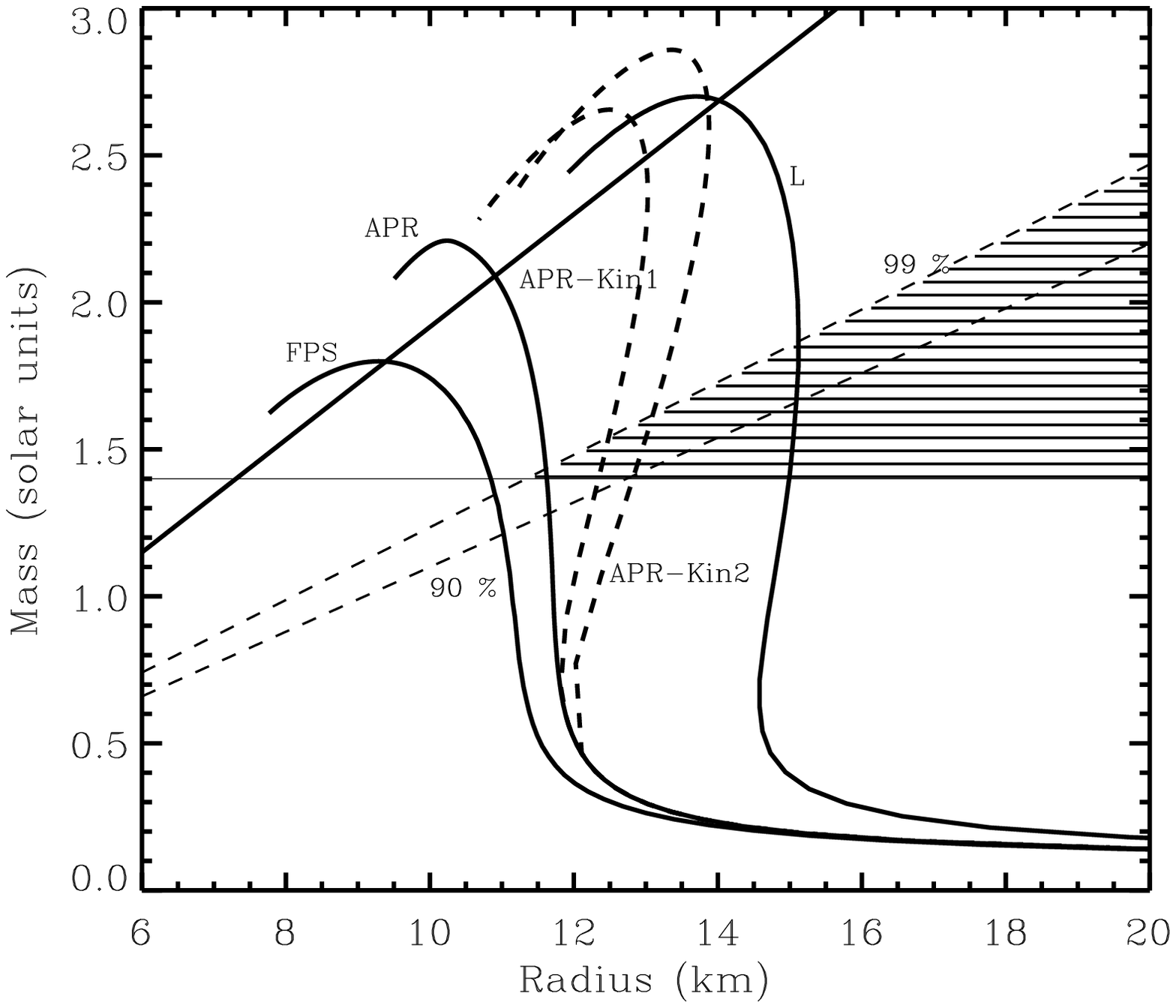]{Summary of mass radius constraints from 
fits to bursts from 4U 1636-53 using the two spot model and the grey atmosphere
beaming function. The diagonal dashed lines show the 90 and 99 \% confidence 
upper limits for $M/R$ from the four fits in Table 1 (see also Figure 5).  The 
shaded region is the allowed range of $M$ and $R$ which satisfies the 
compactness constraints and has $M > 1.4 M_{\odot}$. The solid diagonal line 
corresponds to our computational limit, $M/R = 0.284$.  
The other curves show mass - radius relations for equations of state FPS 
(Lorenz et al. 1993), L (Pandharipande \& Smith 1975b), and APR (Akmal, 
Pandharipande \& Ravenhall 1998), which range from very soft (FPS) to very 
stiff (L). We also show two different modifications to the APR EOS based on the
relativistic kinetic theory constraints of Olson (2001) (thick dashed curves). 
The two curves correspond to the use of different matching densities for the 
high density kinetic theory constraints (see \S 5 and Olson 2001). The results 
favor stiffer equations of state with $R > 11.5$ km for a 1.4 $M_{\odot}$ 
neutron star. \label{fig6}}

\newpage

\begin{figure}
\begin{center}
 \includegraphics[width=6in,height=7.2in]{f1.ps}
\end{center}

Figure 1: Geometry for calculation of the flux from a hot 
spot on a rotating neutron star. Here the hot spot is situated on the 
rotational equator. See the text for a description of the relationship between 
the angles $\phi$, $\psi$ and the impact parameter, $b$.
\end{figure}

\clearpage

\begin{figure}
\begin{center}
 \includegraphics[width=6in, height=7in, angle=90]{f2.ps}
\end{center}

Figure 2: Light-curves generated with the
rotating hot spot model for different values of the neutron star compactness.
Notice the decrease in amplitude with increasing compactness. Note also the 
decrease in amplitude and increase in flux as the hot spot spreads to encompass
the entire surface. These models were computed with one hot spot assuming
isotropic emission from the surface. The top three curves have been displaced
vertically for clarity. The curves were computed using isotropic emission from
the surface, the qualitative behavior of the amplitude with compactness, 
$M/R$, using the grey atmosphere beaming function is the same, only the 
modulation amplitudes differ slightly. 
\end{figure}

\clearpage

\begin{figure}
\begin{center}
 \includegraphics[width=5in, height=6in]{f3.ps}
\end{center}
\vskip 10 pt
Figure 3: Model fits to the bursts from 4U 1636-53. Each
panel shows the data (histogram) and model (thick solid curve) fit to the 
rising portion of a burst. The dashed vertical lines denote the time interval
in which we fit the hot spot model. The extent of each model curve covers the 
total time it takes for the hot spots to envelope the entire neutron star
surface. The bursts are labelled by date.
\end{figure}

\clearpage

\begin{figure}
\begin{center}
 \includegraphics[width=6in, height=6in,angle=90]{f4.ps}
\end{center}

Figure 4: Data and best fit models for several 
fits to the December 28th, 1996 burst from 4U 1636-53. Shown are fits using
two hot spots with $\theta_s = \theta_{obs} = 0$ (solid); one hot spot with
$M/R$ fixed at 0.284 and $\theta_s = \theta_{obs} = 0$ (dashed); and one hot 
spot with $\theta_s$ and $\theta_{obs}$ free to vary (dotted). All the fits
shown were computed with the grey atmosphere beaming function.
\end{figure}

\clearpage

\begin{figure}
\begin{center}
 \includegraphics[width=6in, height=5in]{f5.ps}
\end{center}
\vskip 10 pt

Figure 5: Compactness constraints for the four bursts from 4U 1636-53. The
solid horizontal line is the best fitting constant value of compactness,
$\beta_{avg}$. The dashed and dot-dashed lines are the 90 and $99\%$ 
confidence upper limits on $\beta_{avg}$. The burst number corresponds
to their position in Table 1. 
\end{figure}

\clearpage

\begin{figure}
\begin{center}
 \includegraphics[width=6in, height=5in]{f6.ps}
\end{center}

Figure 6: Summary of mass radius constraints from 
fits to bursts from 4U 1636-53 using the two spot model and the grey atmosphere
beaming function. The diagonal dashed lines show the 90 and 99 \% confidence 
upper limits for $M/R$ from the four fits in Table 1 (see also Figure 5).  The 
shaded region is the allowed range of $M$ and $R$ which satisfies the 
compactness constraints and has $M > 1.4 M_{\odot}$. The solid diagonal line 
corresponds to our computational limit, $M/R = 0.284$.  
The other curves show mass - radius relations for equations of state FPS 
(Lorenz et al. 1993), L (Pandharipande \& Smith 1975b), and APR (Akmal, 
Pandharipande \& Ravenhall 1998), which range from very soft (FPS) to very 
stiff (L). We also show two different modifications to the APR EOS based on the
relativistic kinetic theory constraints of Olson (2001) (thick dashed curves). 
The two curves correspond to the use of different matching densities for the 
high density kinetic theory constraints (see \S 5 and Olson 2001). The results 
favor stiffer equations of state with $R > 11.5$ km for a 1.4 $M_{\odot}$ 
neutron star. 
\end{figure}

\clearpage

\hoffset -55pt
\begin{deluxetable}{cccccccc}
\rotate
\tablecolumns{8}
\tablewidth{0pc}
\tablecaption{Summary of Fits to Burst Oscillations in 4U1636-53 and 
4U 1728-34\tablenotemark{a}}
\tablehead{\colhead{Epoch (UTC)} & \colhead{$\Delta T_{fit}$ (s)} & 
\colhead{$\beta = M/R$} & \colhead{$S$ (cts s$^{-1}$ ster$^{-1}$)
\tablenotemark{c}} & 
\colhead{$\alpha_0$ (deg)} & \colhead{$\dot\alpha$ (deg s$^{-1}$)} & 
\colhead{$\delta_0$\tablenotemark{b}} & \colhead{$\chi^2$} }
\startdata
\cutinhead{4U 1636-53}
12/28/96 at 22:39:34 & 0.276 & $0.134\pm 0.037$ & $1641.6$ & 
$6.41\pm 1.6$ & $222.9\pm 39.6$ & $116.7$ & 53.6 \\
08/19/98 at 11:47:07 & 0.303 & $0.133\pm 0.056$ & $1716.1$ & 
$14.38\pm 3.8$ & $130.7\pm 54.3$ & 81.0 & 63.7 \\
08/20/98 at 05:16:35 & 0.496 & $0.075\pm 0.072$ & 882.1 &
$8.66\pm 3.0$ & $69.5\pm 36.0$ & -3.0 & 65.3 \\
06/18/99 at 23:50:10 & 0.460 & $0.127\pm 0.037$ & 1272.3 & 
$6.70\pm 1.4$ & $102.4\pm 22.2$ & 14.8 & 61.5 \\
\cutinhead{4U 1728-34}
02/16/96 at 10:00:49 & 0.221 & $0.113\pm 0.042$ & 1238.7 & 
$9.1\pm 2.2$ & $370.8\pm 39.6$ & 102.7 & 69.7 \\
09/21/97 at 18:10:56 & 0.354 & $0.130\pm 0.043$ & 1001.9 &
$7.4\pm 2.0$ & $181.8\pm 32.4$ & 142.1 & 54.6 \\
\enddata
\tablenotetext{a}{All fits are for two hot spots assuming the grey 
atmosphere beaming function, and have 59 degrees of freedom.}
\tablenotetext{b}{The initial phase is an azimuthal angle, measured with 
respect to the observers direction, and counted positive in the 
anti-clockwise rotational direction of the neutron star. It can be determined 
to $\lesssim 1\%$.}
\tablenotetext{c}{The normalization $S$ is given in terms of the counting 
rate per unit solid angle of the neutron star covered by the hot spot(s). For
example, the peak counting rate can be determined for a given burst by 
multiplying $S$ by $4\pi$. The uncertainty in $S$ is typically about $30\%$. 
This results from the fact that $S$ is strongly correlated with both 
$\dot\alpha$ and $\beta$.} 
\end{deluxetable}

\end{document}